\documentstyle[prl,aps,preprint,epsf]{revtex}
\begin{document}
\draft

\title{Structural and Magnetic Instabilities of La$_{2-x}$Sr$_x$CaCu$_2$O$_6$}

\author{C.~Ulrich$^1$, S.~Kondo$^2$, M.~Reehuis$^3$, H.~He$^1$,
C. Bernhard$^1$, C. Niedermayer$^4$, F. Bour\'ee$^5$, \\
P. Bourges$^5$, M.~Ohl$^6$, H.M.~R\o{}nnow$^7$, H. Takagi$^2$ and
B.~Keimer$^1$}
\address{$^1$Max-Planck-Institut~f\"{u}r~Festk\"{o}rperforschung, Heisenbergstr. 1, D-70569 Stuttgart, Germany}
\address{$^2$Department of Advanced Materials Science, School of Frontier Sciences,
             University of Tokyo, Hongo 7-3-1, Bunkyo-ku, Tokyo 113-8656, Japan}
\address{$^3$Hahn--Meitner--Institut, Glienicker Str. 100, D-14109 Berlin, Germany}
\address{$^4$Department of Physics, Universit\"{a}t Konstanz, D-78434 Konstanz, Germany}
\address{$^5$Laboratoire Le\'on Brillouin, CEA--CNRS, CE Saclay, 91191 Gif sur Yvette, France}
\address{$^6$Institut Laue--Langevin, 156X, 38042 Grenoble Cedex 9, France}
\address{$^7$CEA (MDN/SPSMS/DRFMC), 38054 Grenoble, France}
\date{\today}

\maketitle

\begin{abstract}
A neutron scattering study of nonsuperconducting
La$_{2-x}$Sr$_x$CaCu$_2$O$_6$ (x=0 and 0.2), a bilayer copper
oxide without CuO chains, has revealed an unexpected
tetragonal-to-orthorhombic transition with a doping dependent
transition temperature. The predominant structural modification
below the transition is an in-plane shift of the apical oxygen. In
the doped sample, the orthorhombic superstructure is strongly
disordered, and a glassy state involving both magnetic and
structural degrees of freedom develops at low temperature. The
spin correlations are commensurate.
\end{abstract}

\pacs{74.25.Jb, 74.25.Ha, 61.2.-q, 74.72.Dn}

\clearpage

The interplay between ordering phenomena of charge, spin and
lattice degrees of freedom has been a focus of recent research on
high temperature superconductors. Most of the work has thus far
concentrated on two systems: La$_{2-x}$Sr$_x$CuO$_4$ (LSCO) and
YBa$_2$Cu$_3$O$_{6+x}$ (YBCO). In both systems, spin freezing
phenomena
\cite{sternlieb90,keimer92,chou95,niedermayer98,curro00,sonier01}
as well as magnetic and structural superlattice formation
\cite{tranquada96,lee99,sidis01,mook01} have been observed. The
relationship between these effects, some of which were attributed
to charge ordering, and anomalies detected by spectroscopic probes
(such as the pseudogap and phonon anomalies) is currently an area
of very active investigation. As only two compounds have been
studied in detail, it has proven difficult to disentangle aspects
that are generic to the cuprates from those that are consequences
of some non-generic element of the lattice structure of a specific
compound. Such elements include the CuO chains in YBCO, and soft
optical phonons associated with tilting of the CuO$_6$ octahedra
in LSCO. For instance, charge carriers in the YBCO chains are
prone to localization and charge order and may induce related
instabilities in the CuO$_2$ planes. Tilt instabilities of the
CuO$_6$ octahedra, on the other hand, are known to be an important
stabilizing factor for static ``stripe" order in LSCO and its
derivatives \cite{tranquada96}.

In order to further explore the universality of the structural and
electronic instabilities in the cuprates, we have carried out a
neutron scattering study of La$_{2-x}$Sr$_x$CaCu$_2$O$_6$ (LSCCO),
a compound that crystallizes in a bilayer structure akin to YBCO,
but without any electronically active element other than the
CuO$_2$ layers (inset in Fig. \ref{fig2})
\cite{cava90,shaked93,deng99}. Further, tilt instabilities of the
kind observed in LSCO are not expected in LSCCO because of the
fivefold, pyramidal coordination of copper. LSCCO has thus been
termed the simplest bilayer cuprate \cite{cava90}. We report the
discovery of an unexpected low temperature
tetragonal-to-orthorhombic transition in undoped and doped LSCCO.
In the doped sample, the orthorhombic superstructure does not
develop long range order, and the magnetic and structural
correlations evolve with temperature in a strikingly parallel
manner. Despite the large nominal hole concentration (0.1 holes
per Cu), the spin correlations are commensurate.

Single crystals of La$_{2-x}$Sr$_x$CaCu$_2$O$_6$ were grown by the
travelling-solvent floating zone method under 1 atm oxygen flow.
Most of the measurements were taken on an x=0.2 single crystal of
volume $3 \times 3 \times 3$~mm$^3$, which is insulating at low
temperature, although a metallic behavior of the resistivity is
observed at high temperature. (For reasons that are poorly
understood at present, superconductivity occurs only in samples
treated under high oxygen pressure.) The neutron experiments were
carried out on the four-circle diffractometer E5 at the BER II
reactor of the Hahn-Meitner-Institut in Berlin, Germany, and on
the triple axis spectrometer IN22 at the Institut Laue-Langevin in
Grenoble, France. The former instrument was used to acquire a
catalogue of Bragg intensities for an accurate refinement of the
crystal structure, while the latter instrument allowed us to
measure the temperature dependences of selected structural and
magnetic reflections with good statistics. On E5, neutron
wavelengths of 0.889~\AA~ or 2.36~\AA~ were selected by a copper
or a pyrolytic graphite (PG) monochromator, respectively, and the
data were collected with a two-dimensional position sensitive
$^3$He-detector. On IN22 the (002) reflection of PG, set for a
neutron wavelength of 2.36~\AA, was used as both monochromator and
analyser, and in order to avoid second order contamination a PG
filter was inserted into the incident beam. In order to optimize
the intensity, the instrument was used in a focusing mode without
collimations.

Two data sets with a total of 623 (259 unique) and 791 (256
unique) reflections were collected on E5 at room temperature and
10~K, respectively. Excellent structure refinements of the room
temperature data could be obtained using the tetragonal space
group $I4/mmm$. The resulting lattice parameters, atomic
coordinates and isotropic temperature factors are listed in Table
\ref{tabcrystal} and are in good agreement with those of previous
reports \cite{cava90,shaked93,deng99}; further details will be
reported elsewhere. At 10 K, however, weak additional (and
hitherto unobserved) reflections were found, as shown in Fig.
\ref{fig1}. Since their intensity increases with increasing
momentum transfer $Q$, these reflections must be attributed to a
structural modulation. Their temperature dependence (Fig.
\ref{fig2}b) shows that the superstructure develops around 100 K,
with diffuse scattering persisting to somewhat higher
temperatures.

The superstructure reflections can be indexed as $(h/2,~k/2,~l)$
with $h$, $k$ odd and $l$ even but $\neq 0$, indicating a doubling
of the unit cell. From these conditions, the orthorhombic space
group $Bmab$ (No. 64, standard setting $Cmca$) with lattice
dimensions $a \textstyle{\sqrt2} \times b \textstyle{\sqrt2}
\times c$ can be deduced. The momentum resolution of our
experiment was not sufficient to resolve the small difference
between $a$ and $b$, and for simplicity we continue to use the
$I4/mmm$ indexing for the remainder of this article. The atomic
coordinates resulting from a refinement of the low temperature
data are also given in Table \ref{tabcrystal}. The primary
difference between the $I4/mmm$ and $Bmab$ structures is a shift
of the apical oxygen along the tetragonal (110) direction by
$\sim$ 0.05~\AA.

The single-layer sister compound LSCO also undergoes a transition
from a high temperature tetragonal ($I4/mmm$) to a low temperature
orthorhombic ($Bmab$) structure \cite{keimer92}, but this
transition occurs at a much higher temperature and is commonly
believed to be due to steric effects: a mismatch between the
natural lattice spacings of the (La/Sr)O and CuO$_2$ layers is
thought to be relieved through a staggered tilt of the CuO$_6$
octahedra (which involves motions of both in-plane and apical
oxygen atoms). In LSCCO, on the other hand, the coordination of
copper is fivefold (inset in Fig. \ref{fig2}) and the displacement
pattern (Table \ref{tabcrystal}) involves almost exclusively the
apical oxygen. The origin of the tetragonal-to-orthorhombic
transition in LSCCO thus merits some consideration, and as a
resistivity minimum occurs for LSCCO samples in this doping range
around the same temperature at which the superstructure
reflections first appear \cite{kinoshita92}, one may wonder
whether the transition could be a consequence of charge
localization.

In order to explore this possibility, we have carried out a
neutron diffraction study of a Sr-free La$_2$CaCu$_2$O$_6$ (LCCO)
powder. Although as-prepared LCCO contains a small density of
excess oxygen, its resistivity is much larger (and the charge
carrier concentration therefore lower) than in LSCCO with x=0.2
\cite{kinoshita92}. The experiment was carried out on the 3T2
powder diffractometer at the Laboratoire L\'eon Brillouin in
Saclay, France, with a neutron wavelength of 1.2252~\AA. While the
room temperature data were consistent with previous work
\cite{cava90,shaked93,deng99}, additional weak structural
reflections obeying the same extinction rules as for the x=0.2
compound were again observed at lower temperature (Fig.
\ref{fig2}a). The displacement of the apical oxygen is
significantly larger than for x=0.2, and the structural transition
temperature ($170 \pm 20$ K) is higher.

These data demonstrate the presence of an intrinsic structural
instability even without charge carriers, and the enhanced
transition temperature in LCCO suggests that the doped carriers
suppress the structural phase transition. This is confirmed by
considering the widths of the corresponding reflections which are
strongly broadened in the x=0.2 sample (Fig. \ref{fig2}c). At low
temperature we infer correlation lengths of about ten lattice
parameters both in plane and out of plane. Further, the
correlation length decreases rapidly as the temperature is
increased, which indicates pronounced low energy fluctuations of
the corresponding order parameter (that is, the apical oxygen
displacement). This result contrasts sharply with the situation in
LSCO: While doping-induced {\it local} structural defects have
been reported \cite{bozin00}, the orthorhombic superstructure
remains long range ordered even in highly doped LSCO. The
structural data of Fig. \ref{fig2}c are in fact reminiscent of the
rapid evolution of the {\it magnetic} correlation length with
doping and temperature in underdoped LSCO \cite{keimer92}.

With this correspondence in mind, we now turn to the magnetic
properties of LSCCO. In addition to the orthorhombic
superstructure reflections discussed above, Fig. \ref{fig1} shows
that some intensity develops at low temperature around the
antiferromagnetic positions ${\bf Q} = (1/2,~1/2,~l)$ with $l$
odd. In contrast to the structural reflections, this intensity is
only observed at low $Q$ and can hence be identified as magnetic.
It also exhibits a sinusoidal intensity modulation as a function
of $l$ that extinguishes the $l=0$ reflection. A similar
modulation is observed in antiferromagnetic YBCO and is known to
arise from antiferromagnetic interlayer exchange within a bilayer
unit \cite{tranquada89}. The integrated intensity of a
structurally forbidden but magnetically allowed reflection, (1/2,
1/2, 3), is shown in Fig. \ref{fig3}b as a function of
temperature. Its onset temperature ($\sim 90$ K) is identical to
the structural transition temperature (Fig. \ref{fig2}b) to within
the experimental error \cite{note}. The magnetic correlation
length is somewhat shorter than that of the orthorhombic
superstructure at low temperatures but shows a similar evolution
with temperature.

As noted above, this behavior of the magnetic correlation length
is characteristic of underdoped LSCO, where the electronic spins
are known to freeze out gradually and a spin glass state develops
at low temperatures
\cite{sternlieb90,keimer92,chou95,niedermayer98,curro00}. Upon
lowering the temperature, manifestations of magnetic ordering thus
appear progressively in neutron scattering and muon spin rotation
($\mu$SR) measurements (which are sensitive to fluctuations on meV
and $\mu$eV energy scales, respectively). In order to complete the
description of the microscopic magnetic properties of LSCCO, we
have carried out zero-field $\mu$SR measurements on the same LSCCO
sample studied by neutrons. The experiment was performed at the
GPS spectrometer at the Paul-Scherrer-Institute, Switzerland;
typical data are shown in the inset to Fig. \ref{fig3}. A gradual
onset of muon spin relaxation by slow electronic spin fluctuations
is observed upon cooling below 100 K, consistent with the onset of
antiferromagnetic diffuse scattering in the neutron experiment
(Fig. \ref{fig3}b). However, a precession signal heralding
magnetic order that is static on the energy scale set by the muon
Larmor frequency only appears at the much lower temperature of 10
K (Fig. \ref{fig3}a). A magnetic anomaly previously found around
10 K in this doping regime \cite{kinoshita92,ansaldo92,felner93}
can hence be identified as a spin glass transition. Although the
temperature scale is somewhat larger as compared to LSCO, our
neutron scattering and $\mu$SR data on LSCCO underscore the
universality of the spin freezing phenomena for the cuprates.

In summary, we have reported a detailed neutron scattering study
of the LSCCO system. The magnetically disordered state we have
uncovered is common to the underdoped cuprates. However, the spin
correlations found in LSCCO are {\it commensurate}, whereas {\it
incommensurate} spin correlations are observed in LSCO at
comparable Sr concentrations \cite{tranquada96,lee99}. This may
either indicate an effective doping level much lower than
suggested by the Sr concentration, or it may be a consequence of
bilayer interactions; indeed, commensurate quasielastic spin
correlations have also been reported in underdoped YBCO
\cite{sidis01,mook01}. Another novel feature of the data presented
here is the strongly disordered orthorhombic superstructure in the
doped sample. While the substitutional disorder on the La/Ca sites
(Table \ref{tabcrystal}) may play some role, the broadening
originating from such factors is expected to be temperature
independent. The parallel temperature evolution of the magnetic
and structural order parameters and correlation lengths suggests
that structural relaxation of doped holes is the predominant
factor inhibiting long range order of both magnetic and lattice
degrees of freedom. The comparison with LSCO is also interesting
in this context: LSCCO undergoes its structural transition at a
temperature that roughly coincides with the localization of the
doped holes, so that structural disorder created by localized
holes can effectively frustrate the development of long range
order. Charge localization in LSCO, on the other hand, takes place
at a temperature at which the orthorhombic superstructure is
already well developed, which may explain why the structural
manifestations of charge localization are more subtle in that
system \cite{bozin00}.

As these considerations show, the full implications of our data
are yet to be clarified. It is already clear, however, that the
characterization of magnetic and lattice superstructures for a
lattice type different from those of LSCO and YBCO constitutes an
important step in an ongoing effort to develop a comprehensive
description of the interplay between spin, charge and lattice
degrees of freedom in the cuprates.

%---------------------------------------------------------------------------------------------

%-----------------------------------------------------------------------------------------------

\newpage
% \widetext
\begin{table}[h]
\caption{\label{tabcrystal} Positional parameters of
La$_{1.8}$Sr$_{0.2}$CaCu$_2$O$_6$ at 10~K and 295~K from
single--crystal neutron diffraction. For the refinements of the
10~K data the occupancies of the atoms were taken from refinements
of the tetragonal structure and were not allowed to vary. La/Sr
and Ca are partially exchanged, and by taking the different
multiplicities of the Wyckoff positions into account we obtain the
formula
[(La/Sr)$_{1.81}$Ca$_{0.19}$][Ca$_{0.83}$(La/Sr)$_{0.17}$]Cu$_2$O$_6$.
Because the ionic radii of La$^{3+}$ and Sr$^{2+}$ are similar, we
expect a La:Sr ratio of 9:1 on each position.} \vspace{0.1in}
% \squeezetable
\begin{tabular}{llllllll}
$Atom$  &        & $x$   & $y$   & $z$   & $B_{iso}$ [\AA$^2$] & $Pop$ \\
\hline at 295~K & $I4/mmm$ &
\multicolumn{6}{l}{$a = b = 3.7974(4)$~\AA, $c = 19.3138(19)$~\AA} \\
\hline
La/Sr   & $4e$   & 0     & 0     & 0.17600(9)  & 0.58(4)  & 0.90(3)   \\
Ca      & $4e$   & 0     & 0     & 0.17600     & 0.58     & 0.10(3)   \\
Ca      & $2a$   & 0     & 0     & 0           & 0.61(8)  & 0.83(3)   \\
La/Sr   & $2a$   & 0     & 0     & 0           & 0.61     & 0.17(3)   \\
Cu      & $4e$   & 0     & 0     & 0.41447(9)  & 0.56(3)  & 1.0       \\
O1      & $8g$   & 0.25  & 0.25  & 0.08203(8)  & 0.82(5)  & 1.000(14) \\
O2      & $4e$   & 0     & 0     & 0.29639(14) & 1.86(9)  & 1.012(17) \\
\hline at 10 K & $Bmab$ &
\multicolumn{6}{l}{$a = b = 5.3621(5)$~\AA, $c = 19.2436(19)$~\AA} \\
\hline
La/Sr   & $8f$   & 0     & -0.0009(3) & 0.17575(11) & 0.42(4) & 0.90 \\
Ca      & $8f$   & 0     & -0.0009    & 0.17575     & 0.42    & 0.10 \\
Ca      & $4a$   & 0     & 0          & 0           & 0.50(9) & 0.83 \\
La/Sr   & $4a$   & 0     & 0          & 0           & 0.50    & 0.17 \\
Cu      & $8f$   & 0     & 0.0002(3)  & 0.41474(11) & 0.47(5) & 1.00 \\
O1      & $8g$   & 0.25  & 0.25       & 0.0831(4)   & 0.65(7) & 1.00 \\
O1'     & $8g$   & 0.25  & 0.25       & 0.9183(4)   & 0.65(7) & 1.00 \\
O2      & $8f$   & 0     & 0.0094(5)  & 0.29634(17) & 1.42(9) & 1.01 \\
\end{tabular}
\end{table}
% \narrowtext

\newpage
\begin{figure}
\begin{minipage}{15cm}
\caption{\label{fig1} Elastic scans along $Q =(0.5, 0.5, l)$ and
$(1.5, 1.5, l)$ at 1.6~K and 150~K in the doped
La$_{1.8}$Sr$_{0.2}$CaCu$_2$O$_6$ single crystal, showing
superstructure reflections at (h/2, k/2, $l$). The wave vector $Q$
is given in reciprocal lattice units (r.l.u.) based on the lattice
parameters of Table \ref{tabcrystal}. The intensity for odd $l$
decreases with increasing $Q$ and is thus of magnetic origin. The
intensity for even values of $l$ increases with increasing $Q$ and
is thus predominantly of structural origin. Diffraction peaks
originating from the Al sample holder are marked.}
\end{minipage}
\end{figure}

%---------------------------------------------------------------------------------------------

\begin{figure}
\epsfxsize=10cm
\begin{minipage}{15cm}
\caption{\label{fig2} a) Integrated intensity of the orthorhombic
superstructure reflection (3/2,~3/2,~4) of the undoped
La$_2$CaCu$_2$O$_6$ powder sample and b) of the doped
La$_{1.8}$Sr$_{0.2}$CaCu$_2$O$_6$ single crystal as a function of
temperature. Panel c shows the temperature dependence of the
intrinsic full width at half maximum (FWHM) of the (3/2, 3/2, 4)
peak measured along $(h, h, 0)$ in the doped compound. The lines
are guides-to-the-eye. The inset displays the unit cell at T = 295
K. The $\rm CuO_5$ pyramids are highlighted. (The Cu atoms at the
center of the pyramid base are not shown for clarity.)}
\end{minipage}
\end{figure}

%---------------------------------------------------------------------------------------------

\begin{figure}
\begin{minipage}{15cm}
\caption{\label{fig3} a) Temperature dependence of the zero-field
$\mu$SR Lamor frequencies measured in the
La$_{1.8}$Sr$_{0.2}$CaCu$_2$O$_6$ single crystal. The two distinct
frequencies indicate two inequivalent muon sites. The inset
displays the asymmetry versus time spectrum for two different
temperatures above and below the magnetic ordering temperature.
The lines are guides-to-the-eye. b) Temperature dependence of the
integrated intensity of the (1/2, 1/2, 3) reflection measured by
neutron scattering.}
\end{minipage}
\end{figure}


\newpage
\begin{references}
\bibitem{sternlieb90} B.J. Sternlieb, G.M. Luke, Y.J. Uemura, T.M. Riseman,
J.H. Brewer, P.M. Gehring, K. Yamada, Y. Hidaka and T. Murakami,
T.R. Thurston, and R.J. Birgeneau, Phys. Rev. B {\bf 41}, 8866
(1990).
\bibitem{keimer92} B. Keimer, N. Belk, R.J. Birgeneau, A. Cassanho, C.Y. Chen, M. Greven,
M.A. Kastner, A. Aharony, Y. Endoh, R.W. Erwin, and G. Shirane,
Phys. Rev. B {\bf 46}, 14034 (1992).
\bibitem{chou95} F.C. Chou, N.R. Belk, M.A. Kastner, R.J. Birgeneau, and A. Aharony,
Phys. Rev. Lett. {\bf 75}, 2204 (1995).
\bibitem{niedermayer98} C. Niedermayer, C. Bernhard, T. Blasius, A. Golnik,
A. Moodenbaugh, and J.I. Budnick, Phys. Rev. Lett. {\bf 80}, 3843 (1998).
\bibitem{curro00} N.J. Curro, P.C. Hammel, B.J. Suh, M. H\"{u}cker, B. B\"{u}chner, U.
Ammerahl, and A. Revcolevschi, Phys. Rev. Lett. {\bf 85}, 642
(2000).
\bibitem{sonier01} J.E. Sonier, J.H. Brewer, R.F. Kiefl, R.I. Miller, G.D. Morris, C.E. Stronach, J.S.
Gardner, S.R. Dunsiger, D.A. Bonn, W.N. Hardy, R. Liang, and R.H.
Heffner, Science {\bf 292}, 1692 (2001).
\bibitem{tranquada96} J.M. Tranquada, N. Ichikawa, and S. Uchida, Phys. Rev. B {\bf 59}, 14712 (1999).
\bibitem{lee99} Y.S. Lee, R.J. Birgeneau, M.A. Kastner, Y. Endoh, S. Wakimoto, K. Yamada,
R.W. Erwin, S.H. Lee, and G. Shirane, Phys. Rev. B {\bf 60}, 3643
(1999).
\bibitem{sidis01} Y. Sidis, C. Ulrich, P. Bourges, C. Bernhard, C.  Niedermayer, L.P. Regnault, N.H. Andersen, and
B. Keimer, Phys. Rev. Lett. {\bf 86}, 4100 (2001).
\bibitem{mook01} H.A. Mook, P. Dai, and F. Dogan, Phys. Rev. B {\bf 64}, 012502
(2001).
\bibitem{cava90} R.J. Cava, B. Batlogg, R.B. Vandover, J.J. Krajewski, J.V. Waszczak, R.M. Fleming, W.F. Peck, L.W. Rupp,
P. Marsh, A.C.W.P. James, and L.F. Schneemeyer, Nature {\bf 345},
602 (1990).
\bibitem{shaked93} H. Shaked, J.D. Jorgensen, B.A. Hunter, R.L. Hitterman, K. Kinoshita, F. Izumi, and T. Kamiyama,
Phys. Rev. B {\bf 48}, 12941 (1993).
\bibitem{deng99} H. Deng, C. Dong, H. Chen, F. Wu, S.L. Jia, J.C. Shen, and Z. X. Zhao, Physica C {\bf 313}, 285 (1999).
\bibitem{kinoshita92} K. Kinoshita and T. Yamada, Phys. Rev. B {\bf 46}, 9116 (1992).
\bibitem{bozin00} E.S. Bozin, G.H. Kwei, H. Takagi and S.J.L.
Billinge, Phys. Rev. Lett. {\bf 84}, 5856 (2000).
\bibitem{tranquada89} J.M. Tranquada, G. Shirane, B. Keimer, S. Shamoto and M. Sato, Phys. Rev. B {\bf 40}, 4503
(1989).
\bibitem{note} The (3/2, 3/2, 4) reflection shown in Fig. \ref{fig2}b is both
structurally and magnetically allowed, but only about 5~\% of its
intensity is of magnetic origin.
\bibitem{ansaldo92} E.J. Ansaldo, C. Niedermayer, H. Gl\"{u}ckler, C.E. Stronach, T.M. Riseman, D.R. Noakes,
X. Obradors, A. Fuertes, J.M. Navarro, P. Gomez, N. Casan, B.
Martinez, F. Perez, J. Rodriguez-Carvajal, R.S. Cary, and K. Chow,
Phys. Rev. B {\bf 46}, 3084 (1992).
\bibitem{felner93} I. Felner, D. Hechel, E.R. Yacoby, G. Hilscher, T. Holubar, and G. Schaudy, Phys. Rev. B {\bf 47},
12190 (1993).
\end{references}
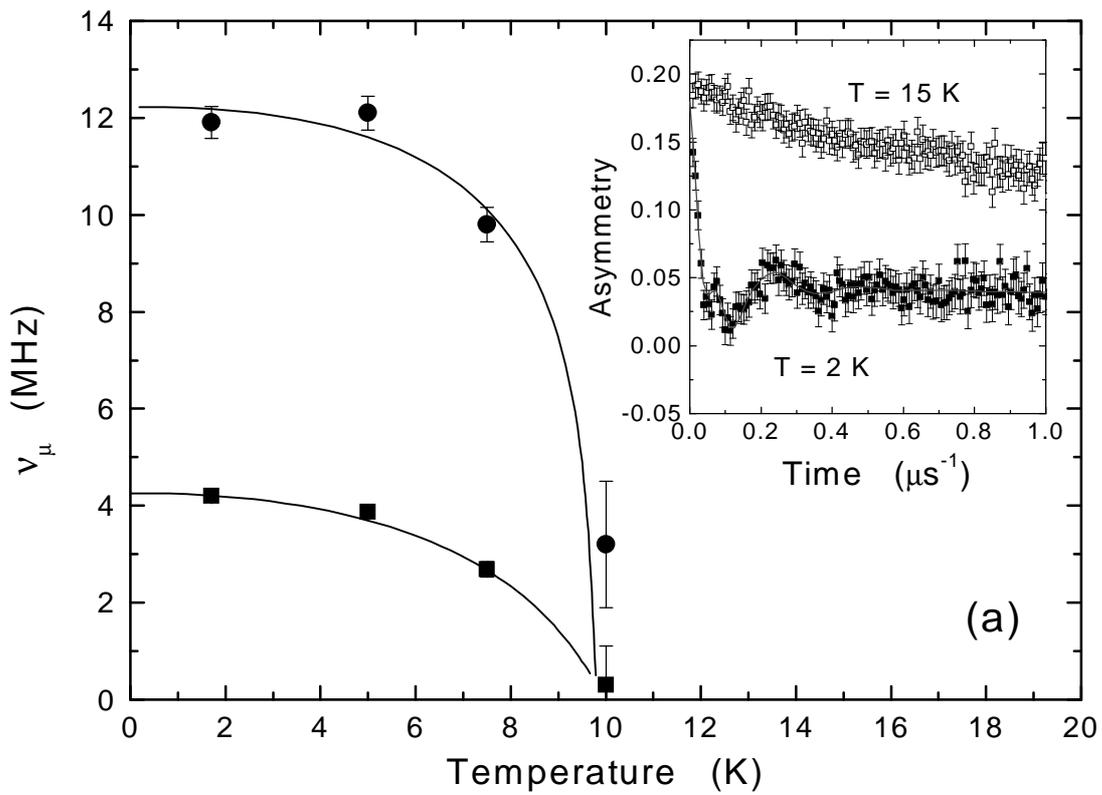
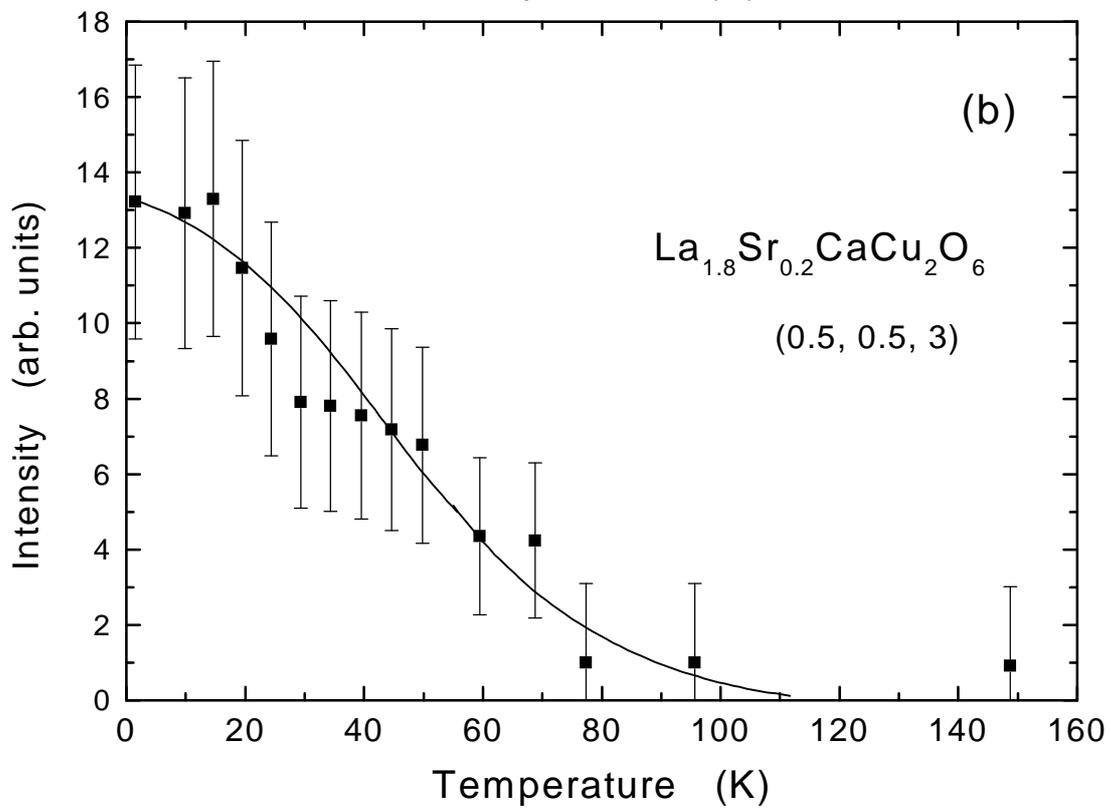
\end{document}